\documentclass[journal]{IEEEtran}
\usepackage[latin9]{inputenc}
\usepackage{amsmath}
\usepackage{amsthm}
\usepackage{amssymb}
\usepackage{graphicx}
\usepackage{wasysym}
\usepackage[unicode=true,
 bookmarks=true,bookmarksnumbered=true,bookmarksopen=true,bookmarksopenlevel=1,
 breaklinks=false,pdfborder={0 0 0},pdfborderstyle={},backref=false,colorlinks=false]
 {hyperref}
\hypersetup{
 pdfborderstyle={}}

\makeatletter
\theoremstyle{plain}
\newtheorem{thm}{\protect\theoremname}
\theoremstyle{definition}
\newtheorem{defn}[thm]{\protect\definitionname}
\theoremstyle{remark}
\newtheorem{rem}[thm]{\protect\remarkname}

\usepackage{amsthm}\usepackage{wasysym}
\usepackage{cite}
\theoremstyle{plain}
\theoremstyle{definition}

\@ifundefined{showcaptionsetup}{}{%
 \PassOptionsToPackage{caption=false}{subfig}}
\usepackage{subfig}
\makeatother

\providecommand{\definitionname}{Definition}
\providecommand{\remarkname}{Remark}
\providecommand{\theoremname}{Theorem}

\begin{document}

\title{Wi-Fi Coexistence with Duty Cycled LTE-U}

\author{Yimin Pang, Alireza Babaei, Jennifer Andreoli-Fang and Belal Hamzeh
\thanks{Y. Pang is with Department of Electrical, Computer and Energy Engineering,
University of Colorado Boulder, Boulder, CO 80309, Email: yipa5803@colorado.edu} \thanks{A. Babaei is currently with Ofinno Technologies, Herndon, VA , Email:
ababaei@ieee.org} \thanks{J. Andreoli-Fang and B. Hamzeh are currently with Cable Television
Laboratories, Louisville, CO, Email: \{j.fang, b.hamzeh\}@cablelabs.com}}
\maketitle
\begin{abstract}
Coexistence of Wi-Fi and LTE-Unlicensed (LTE-U) technologies has drawn
significant concern in industry. In this paper, we investigate the
Wi-Fi performance in the presence of duty cycle based LTE-U transmission
on the same channel. More specifically, one LTE-U cell and one Wi-Fi
basic service set (BSS) coexist by allowing LTE-U devices transmit
their signals only in predetermined duty cycles. Wi-Fi stations, on
the other hand, simply contend the shared channel using the distributed
coordination function (DCF) protocol without cooperation with the
LTE-U system or prior knowledge about the duty cycle period or duty
cycle of LTE-U transmission. We define the fairness of the above scheme
as the difference between Wi-Fi performance loss ratio (considering
a defined reference performance) and the LTE-U duty cycle (or function
of LTE-U duty cycle). Depending on the interference to noise ratio
(INR) being above or below -62dbm, we classify the LTE-U interference
as strong or weak and establish mathematical models accordingly. The
average throughput and average service time of Wi-Fi are both formulated
as functions of Wi-Fi and LTE-U system parameters using probability
theory. Lastly, we use the Monte Carlo analysis to demonstrate the
fairness of Wi-Fi and LTE-U air time sharing. 
\end{abstract}

\begin{IEEEkeywords}
coexistence, LAA-LTE, LTE-U, LTE unlicensed, medium access delay,
service time, throughput, Wi-Fi, WLAN 
\end{IEEEkeywords}

\section{Introduction}

\IEEEPARstart{T}{he} rapidly growing demand of wireless network services
has led the mobile network operators (MNOs) to look into the possibility
of exploring unlicensed spectrum to offload the data traffic from
the licensed bands. Most recently, 3GPP and other industry alliances
are considering extending LTE into the unlicensed spectrum\footnote{3GPP has completed standardization of licensde assisted access (LAA)
in release 13 (DL only) and enhanced LAA (eLAA) standardization, where
UL access to unlicensed spectrum is also considered, is currently
onoing in 3PP release 14. LAA and eLAA incorporate listen-before-talk
(LBT) for their channel access to the unlicensed spectrum. In addition,
the LTE-U forum, an industry alliance formed by some vendors and mobile
operators, has released an LTE supplemental downlink (SDL) coexistence
specification, where an adaptive duty cycle based coexistence scheme
is introduced. The term ``LTE-U'' is used instead of ``LAA'' in
the LTE-U forum specification. Since this paper focuses on duty cycle
based LTE on unlicensed spectrum, we use the short term LTE-U for
convenience. However, the channel access approach introduced in this
paper is not completely aligned with the channel access method introduced
by the LTE-U forum specification due to non-adaptive duty cycle.}, to offload part of the LTE data traffic onto the unlicensed spectrum.
Compared to data offload using Wi-Fi, this approach has the advantage
of seamless integration into the existing LTE evolved packet core
(EPC) architecture. In a proposal outlined in \cite{qualcommlteuwhitepaper},
three LTE-U modes are introduced as supplement downlink, TD-LTE-U
carrier aggregation and standalone, the first two of which were proposed
to 3GPP as possible candidates.

\begin{figure*}[ht]
\begin{centering}
\includegraphics[scale=0.6]{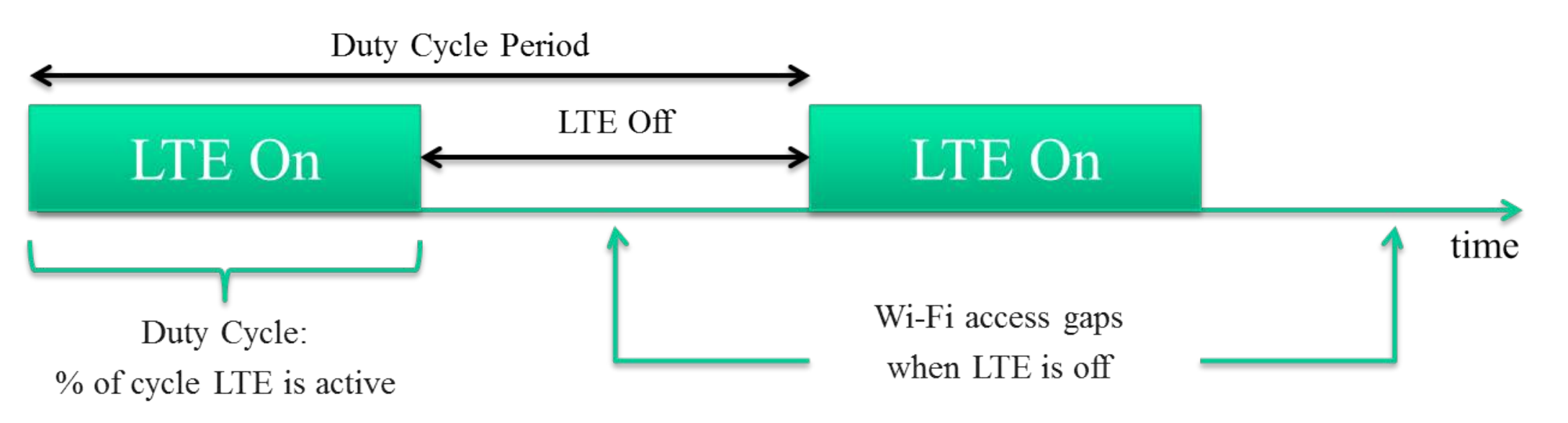} 
\par\end{centering}
\caption{Air time sharing LTE-U\label{fig:Duty-cycle-LTE}}
\end{figure*}

Coexistence of heterogeneous networks such as Wi-Fi and LTE-U on the
same band and their uncoordinated operations can potentially cause
significant interference, degrading the performance of both systems.
Solutions to manage the interference between such systems are therefore
necessary for their successful coexistence. One straightforward method
is to split the common radio channel through air time sharing between
the Wi-Fi and LTE-U sub-systems. With this approach, LTE-U operates
over the shared channel periodically\footnote{More specifically, the radio resources of LTE-U that reside in the
unlicensed spectrum and are shared with Wi-Fi are utilized periodically.}, and during each period (the so-called duty cycle period, denoted
as $T$), only a portion (defined by the LTE-U duty cycle $\alpha$)
of the time is utilized for the LTE-U transmission, as shown in Fig.\,\ref{fig:Duty-cycle-LTE}.
In this scheme, Wi-Fi has no cooperation with LTE-U. Wi-Fi stations
have neither knowledge about the time length of duty cycle period
nor the duty cycle, they simply access the shared channel by standard
channel sensing and random back-off mechanisms.

In an LTE system, each of the user equipments (UEs) communicates with
a base station (eNB) in a deterministic manner through a centralized
channel access control mechanism. The access time and OFDM sub-carriers
of an LTE frame are predetermined at eNB, where the MAC scheduler
considers the radio measurement and quality of service needed for
each UE in its scheduling decisions. Given the above time sharing
scheme and the LTE's centralized access structure, computation of
the LTE-U performance in terms of throughput and service time is relatively
simple and straightforward. However, the medium access mechanism used
by Wi-Fi, controlled by distributed coordination function (DCF) protocol
defined in IEEE 802.11 standard, is random and distributed. Therefore,
we focus on the impact of LTE-U interference on Wi-Fi performance
in this paper.

\subsection{Previous works}

LTE-U and Wi-Fi coexistence is a relatively new area of research.
Previous works in this area are summarized as follows: In \cite{cavalcante2013performance},
Wi-Fi and LTE-U coexistence in single floor and multiple floors environment
at various densities are simulated. The results show that without
any interference management scheme, LTE-U system performance is slightly
affected from Wi-Fi, whereas Wi-Fi is significantly impacted by the
LTE-U transmissions. This result is reinforced in \cite{Babaei2014impact}
by computing the Wi-Fi successful channel accessing probability in
the presence of LTE-U transmission. However, in \cite{qualcommlteuwhitepaper},
LTE-U is described as a better neighbor to Wi-Fi than Wi-Fi to itself
as long as a proper coexistence mechanism (called CSAT) is applied.
Authors in \cite{huaweilteuwhitepaper} present simulation results
on spectrum efficiency comparison between Wi-Fi and LTE-U in a sparse
deployment scenario. The paper, however, lacks sufficient details
on the coexistence features and their effectiveness. Cano and Leith
\cite{cano2015coexistence} proposed a duty-cycle mechanism for LTE-U,
which, by selecting an appropriate probability to access the channel
and transmission duration, ensures proportional fairness among LTE-U
and Wi-Fi nodes. Specifications regarding duty-cycled based LTE-U
are released and maintained by LTE-U forum \cite{lteuforum2015lteu},
where CSAT is officially introduced as the access mechanism.

On the other hand, the 3GPP study item technical report document \cite{3GPPTR36889}
has listed Listen-Before-Talk (LBT) as the required function for clear
channel assessment for LTE LAA. The application of LBT may potentially
enhance the coexistence behavior of Wi-Fi and LTE. Some analysis and
performance test have been reported in \cite{ratasuk2014lte,voicu2015coexistence,kwan2015fair}.

\subsection{Main results}

In this paper, we define Wi-Fi average saturation throughput $\mathcal{R}(T,\alpha,\mathcal{H})$
(in terms of bits/Wi-Fi slot time) and average service time $\mathcal{D}(T,\alpha,\mathcal{H})$
(in terms of Wi-Fi time slots) to be functions of $T$, $\alpha$
and $\mathcal{H}=\{q,L,n\}$, where the set $\mathcal{H}=\{q,L,n\}$
represents an $n$-clients Wi-Fi sub-system with LTE-U to Wi-Fi collision
probability (Wi-Fi transmission failure probability due to LTE-U transmission)
$q$ and Wi-Fi data payload length $L$ (the length of MAC data payload,
in terms of bytes). Given $\mathcal{H}$, the throughput fairness
(cf. Def.\,\ref{def:Def-Fairness}) of a $(T,\alpha)$ air time sharing
scheme is measured by the difference between average Wi-Fi saturation
throughput loss ratio (with respect to the corresponding non-LTE-U
duty cycle scenario $(\infty,0,\mathcal{H})$ performance) and LTE-U
duty cycle $\alpha$, i.e. $\frac{\mathcal{R}(\infty,0,\mathcal{H})-\mathcal{R}(T,\alpha,\mathcal{H})}{\mathcal{R}(\infty,0,\mathcal{H})}-\alpha$.
In a similar way, the average service time fairness is defined as
$\frac{\mathcal{D}(T,\alpha,\mathcal{H})-\mathcal{D}(\infty,0,\mathcal{H})}{\mathcal{D}(\infty,0,\mathcal{H})}-\frac{\alpha}{1-\alpha}$.
These two fairness measures indicate whether Wi-Fi will lose less
or more than $\alpha$ portion of its performance (in the absence
of LTE-U transmission) if $\alpha$ portion of the channel resource
is shared with LTE-U.

Our first step is to analytically formulate $\mathcal{R}(T,\alpha,\mathcal{H})$
and $\mathcal{D}(T,\alpha,\mathcal{H})$ using a probabilistic framework.
The following two key techniques are employed:

\paragraph{Only one labeled client station among the $n$ Wi-Fi stations being
affected by LTE-U interference}

As first introduced by \cite{bianchi2000performance}, Wi-Fi DCF can
be formulated into a Markov chain model, which was generalized later
in \cite{zhai2004performance} and \cite{banchs2006end}. But when
LTE-U is considered, the Markov property no longer holds, because
of the fact that when LTE-U is off, the Wi-Fi transmission failure
probability is only Wi-Fi to Wi-Fi collision probability; when LTE-U
is on, the Wi-Fi transmission failure probability depends on both
Wi-Fi to Wi-Fi and LTE-U to Wi-Fi collision probability. For this
issue, we make an assumption that only one client station among the
$n$ stations, labeled as Sta-A, is affected by the LTE-U interference.
The other $n-1$ stations render the Wi-Fi background traffic for
the labeled station. When $n$ is chosen to be large enough, the background
traffic could still be approximately modeled using the existing framework
in \cite{bianchi2000performance,zhai2004performance,banchs2006end}.
Under this assumption, functions $\mathcal{R}(T,\alpha,\mathcal{H})$
and $\mathcal{D}(T,\alpha,\mathcal{H})$ are with respect to Sta-A.

\paragraph{Different interference levels lead to different formulations}

The Wi-Fi DCF employs CSMA/CA with binary exponential back-off algorithm.
Depending on the energy level being detected, the back-off timer may
or may not be frozen. In short, an LTE-U transmission with interference
to noise ratio (INR) greater than -62dbm or a neighbor Wi-Fi station
transmission with INR greater than -82dbm will cause the interfered
Wi-Fi station freeze its back-off timer. We refer to \textit{weak
interference} as the LTE-U interference with its INR being less than
-62dbm, and\textit{ strong interference} as interference with INR
greater than -62dbm. The mathematical formulations as well as the
performance results are quite different between the cases of weak
and strong LTE-U interference.

Other assumptions are just inherited from the existing framework by
\cite{bianchi2000performance,zhai2004performance,banchs2006end} on
Wi-Fi DCF: 1) A transmission from one Wi-Fi station can be heard by
all the other $n-1$ Wi-Fi stations, and Wi-Fi to Wi-Fi INR is always
greater than -82dbm; 2) Collisions between Wi-Fis or LTE-U to Wi-Fi
are the only causes to Wi-Fi failure transmission.

Then, we analyze the performance as well as the fairness numerically.
Both the analytical functions built for the weak and strong LTE-U
interference are computationally inefficient and characterizing the
two performance functions in closed form is hard. On the other hand,
implementing Monte Carlo analysis based on these two functions is
simple. It is also difficult and meaningless to show the fairness
over all possible combinations of $T$, $\alpha$ and $\mathcal{H}$.
We focus our attention on the cases when LTE-U to Wi-Fi collision
probability $q=1$, which has wide measure over real systems where
LTE-U INR and Wi-Fi SNR are comparable. Other parameters are also
selected in a reasonable range according to practical system setting.
The results demonstrated in Section IV, support the conclusions below: 
\begin{enumerate}
\item Fix $q=1$: Under strong interference, air time sharing scheme could
approximately achieve the fairness for some $(T,\alpha)$; Under weak
interference, air time sharing scheme is generally unfair; 
\item The fairness measure degrades almost linearly when LTE-U to Wi-Fi
collision probability $q$ increases; 
\item The fairness measure degrades almost linearly when Wi-Fi payload length
$L$ increases. 
\end{enumerate}
The rest of the paper is organized as follows: Section II presents
the Wi-Fi and LTE-U coexistence model and formulates the problem;
Section III characterizes the average Wi-Fi saturation throughput
and average service time in the presence of LTE-U duty cycle; The
impact of duty cycled LTE-U interference to Wi-Fi is discussed in
Section IV; Section V concludes the paper.

\section{System Model and Problem Formulation}

In this section, we first introduce the generalized Markov chain model
for Wi-Fi DCF, then we formulate the Wi-Fi and duty cycled LTE-U coexistence,
lastly we define the fairness measure.

\subsection{Generalized Markov chain model for Wi-Fi DCF}

\begin{figure}
\begin{centering}
\includegraphics[scale=0.55]{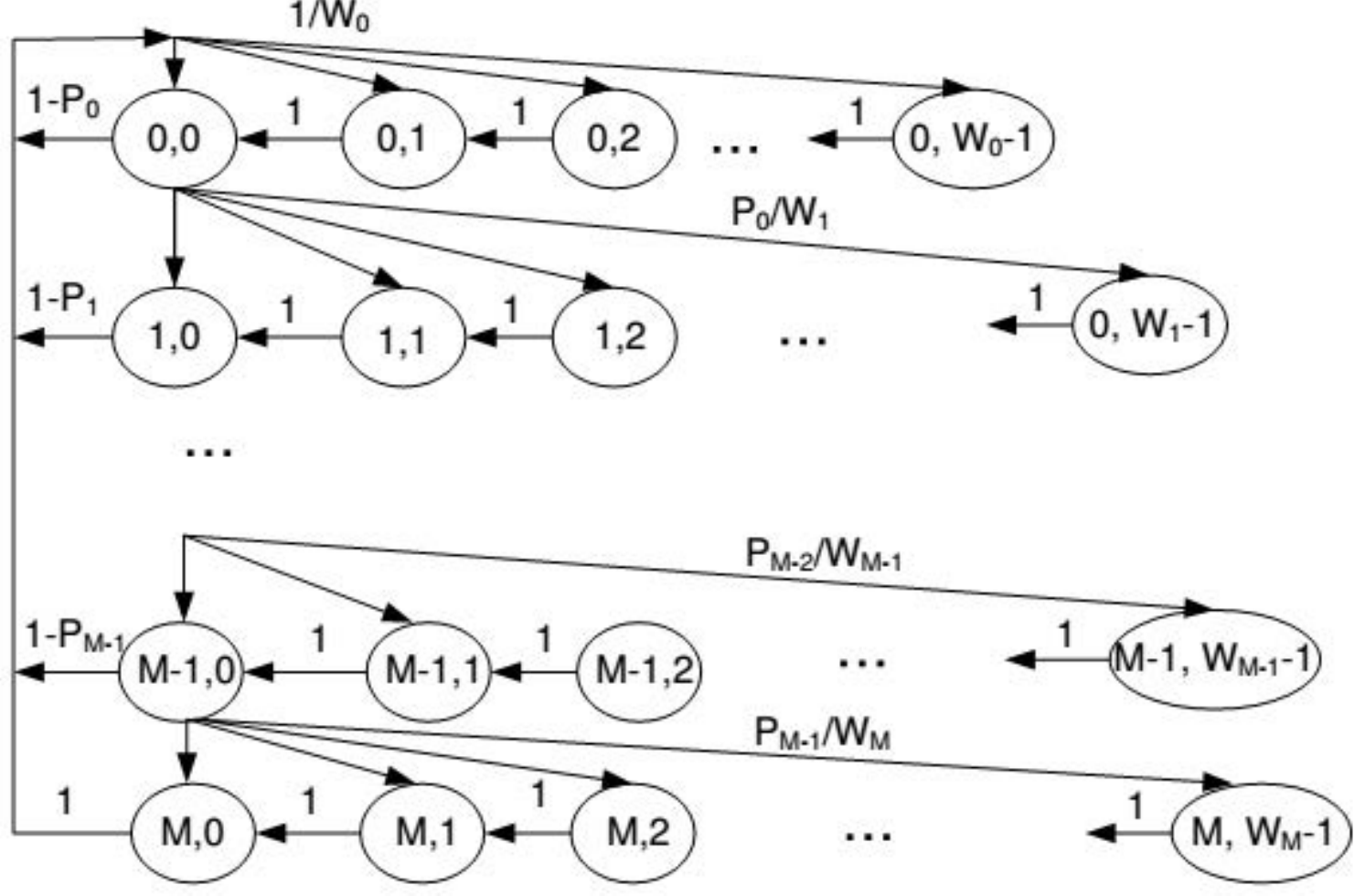} 
\par\end{centering}
\caption{Markov model for Wi-Fi DCF (refer to \cite{bianchi2000performance})\label{fig:MarkovChain}}
\end{figure}

In \cite{bianchi2000performance}, the Wi-Fi DCF is formulated into
a two dimensional Markov chain, the $i$-th floor in the Markov chain
(refer to Fig.\,\ref{fig:MarkovChain}) stands for the random back-off
process before the $i$-th transmission attempt, where $0\leq i\leq M$,
with contention window size $\text{CW}_{i}=2^{i}\text{CW}_{0}$, where
$\text{CW}_{0}$ is the contention window size of the 0-th back-off.
This Markov chain has transition probability $p(i_{n+1},j_{n+1}|i_{n},j_{n})$
(With a slight abuse of notation, we temporarily use $n$ to denote
the state at $n$-th discrete moment),

\begin{equation}
\begin{cases}
1 & i_{n+1}=i_{n};j_{n+1}=j_{n}-1\\
 & j_{n}\neq0\\
1-p_{i_{n}} & i_{n+1}=0,i_{n}\neq M-1;\\
 & j_{n+1}\in\{0,\cdots,\text{CW}_{0}\},j_{n}=0\\
\frac{p_{i_{n}}}{2^{i_{n+1}}\text{CW}_{0}} & i_{n+1}=i_{n}+1,i_{n+1}\neq M-1;\\
 & j_{n+1}\in\{0,\cdots,2^{i_{n+1}}\text{CW}_{0}\},j_{n}=0\\
1 & i_{n+1}=0,i_{n}=M-1;\\
 & j_{n+1}\in\{0,\cdots\text{CW}_{0}\},j_{n}=0
\end{cases}\label{eq:MarkovChain}
\end{equation}

Let $\sigma$ be the duration of the Wi-Fi system slot time as defined
in IEEE 802.11 standard. Throughout the paper, we normalize all the
time variables to $\sigma$, which means 1s is normalized to $1/\sigma$.
During each $(i,0)$ state, a Wi-Fi station senses the channel, with
probability $p_{i}$ it detects clear channel and transmits (or re-transmits)
a packet. If successful, the station stays idle or goes back to 0-th
contention level for a new packet, otherwise the failed packet will
be re-transmitted until it reaches the maximum number of retry attempts
$M$. Before the $i$-th attempt, a random number is generated according
to the uniform distribution $\text{Unif}(0,\text{CW}_{i}-1)$ and
loaded into the back-off timer. The timer decreases the registered
value by one per slot time, once the back-off timer being reset, the
station senses the channel for the $i$-th attempt.

Recall that a Wi-Fi station receiving Wi-Fi interference over -82dbm
will freeze its back-off timer, therefore the transition time between
two neighbor states in the Markov chain may be more than one Wi-Fi
system slot time. In order to use this Markov chain model to analyze
the Wi-Fi service time, the Bianchi model in \cite{bianchi2000performance}
is further generalized by \cite{zhai2004performance} and \cite{banchs2006end}
after incorporating the following two further assumptions: 
\begin{enumerate}
\item The transition time $T_{d}$ between any two neighbor states (we call
this unit decrement time for short, normalized to system slot time)
are identically and independently distributed, and interference between
any two Wi-Fi stations are above -82dbm threshold; 
\item Based on assumption 1 and applying central limit theorem, the time
interval from a back-off timer loads with an initial number $J_{i}$
before $i$-th attempt to the timer being reset is a Gaussian random
variable with mean $J_{i}$E$(T_{d})$. 
\end{enumerate}
We adopt the generalized Markov model of Wi-Fi in this paper.

Next we describe the failure probability $p_{i}$ in each re-transmission
trial. Let $\lambda$ be the probability that there is no packet ready
to transmit, and $\tau$ be the probability that a Wi-Fi station transmits
(or re-transmits) a packet in a randomly chosen time slot given a
packet just left the buffer and is ready to be transmitted. The number
$\tau$ is a function of the number of Wi-Fi stations $n$ and $p_{i}$.
Without LTE-U interference, the probability $p_{i}$ is simply the
collision probability $p_{c}$ that at least two Wi-Fi stations transmits
simultaneously, which is 
\begin{equation}
p_{i}=p_{c}=1-[1-(1-\lambda)\tau]^{n-1}\label{eq:WWCol}
\end{equation}
On the other hand, we have 
\begin{equation}
\tau=\sum_{i=0}^{M-1}(1-p_{c})p(i,0)+(1-p_{c})p(M,0)\label{eq:Tau}
\end{equation}
according to the transition probability defined in \eqref{eq:MarkovChain},
where $p(i,j)$ is the stationary distribution of the Markov chain.
There is no close form expression of the solution to $p_{i}$ and
$\tau$, but given the system parameters and number of stations, they
can be numerically computed. When the Wi-Fi system is saturated, i.e.
the buffer in each station is never empty, i.e. $\lambda=0$ and $p_{c}=1-(1-\tau)^{n-1}$.

It remains to specify the distribution of unit decrement time $T_{d}$.
Let $T_{s}$ and $T_{c}$ be the time duration, normalized to the
system slot time, of one successful and failed (collided) transmission,
respectively. If CTS/RTS mechanism is used, $T_{s}$ and $T_{c}$
can be calculated as follows 
\begin{equation}
T_{s}=\text{RTS}+\text{CTS}+\text{HDR}+L+\text{ACK}+3\times\text{SIFS}+\text{DIFS}\label{eq:TsCtsRts}
\end{equation}
\begin{equation}
T_{c}=\text{RTS}+\text{DIFS}\label{eq:TcCtsRts}
\end{equation}
otherwise 
\begin{equation}
T_{s}=\text{HDR}+L+\text{ACK}+\text{SIFS}+\text{DIFS}\label{eq:TsBasic}
\end{equation}
\begin{equation}
T_{c}=\text{HDR}+L+\text{DIFS}\label{eq:TcBasic}
\end{equation}
In both cases, $L$ denotes the length of the data payload of a Wi-Fi
frame. Let $p_{s}$ be the probability that one of the other $n-1$
Wi-Fi station transmit successfully\footnote{Note it is generally not true that $p_{s}=1-p_{c}$.},
i.e. 
\begin{align}
p_{s} & =(n-1)\tau(1-\tau)^{n-2}\nonumber \\
 & =(n-1)[(1-p_{c})^{\frac{n-2}{n-1}}+p_{c}-1]\label{eq:Psuc}
\end{align}
The unit decrement time $T_{d}$ has following pmf, 
\begin{equation}
p_{T_{d}}(t_{d})=\begin{cases}
1-p_{c} & t_{d}=1\\
p_{c}-p_{s} & t_{d}=T_{c}\\
p_{s} & t_{d}=T_{s}\\
0 & \text{o.w.}
\end{cases}\label{eq:pTd}
\end{equation}

\subsection{Formulation of Wi-Fi and duty cycled LTE-U coexistence}

Consider an infrastructure-based Wi-Fi network coexisting with an
LTE-U network on the same unlicensed band, where interference is coming
from LTE-U sub-system to the Wi-Fi station labeled Sta-A. Considering
a duty cycle period which extends $T$ Wi-Fi system slots (refer to
Fig.\,\ref{fig:Duty-cycle-LTE}) the eNB or UEs in LTE-U sub-system
transmit during the LTE-U ON stage of duration $\alpha T$, where
$\alpha\in[0,1]$, and keep silence during the LTE-U OFF stage\footnote{As discussed in \cite{Babaei2014impact}, even during LTE-U quiet
period, the reference signal may have same significant interference
to Wi-Fi transmission. In this paper, we assume the LTE-U being completely
off during the its OFF stage.}. The variables $T$ and $\alpha$ are defined to be LTE-U \textit{duty
cycle period} and\textit{ duty cycle}, respectively. The Wi-Fi sub-system
does not cooperate with LTE-U nor has any prior knowledge about LTE-U
interference, it simply transmits data frame based on the DCF mechanism.

As has been introduced before, assuming only one out of the $n$ Wi-Fi
stations receive LTE-U interference is for the purpose of maintaining
the Markov properties to model the rest of $n-1$ stations, so when
$n$ is large enough, the collision probability $p_{i}$ and unit
decrement time $T_{d}$ can still be approximately computed using
the generalized Markov chain model. The $n-1$ non-interference stations
actually provide a stationary background Wi-Fi traffic for Sta-A.

We denote a Wi-Fi sub-system as $\mathcal{H}(q,L,n)$, where $q\in[0,1]$
is the Wi-Fi collision probability subject to LTE-U interference (the
Wi-Fi transmission failure probability when a Wi-Fi frame and an LTE-U
frame transmits simultaneously, it only applies to Sta-A), $L$ is
the data payload length in each transmission, and $n$ the number
of clients in the Wi-Fi sub-system which determines the Wi-Fi to Wi-Fi
collision probability (i.e., Wi-Fi collision probability in the absence
of LTE-U transmission). Furthermore let $(T,\alpha)$ denote an air
time sharing scheme with duty cycle period $T$ and LTE-U duty cycle
$\alpha$.

Instead of adopting the uniformed Wi-Fi throughput as in \cite{bianchi2000performance},
we evaluate the Wi-Fi throughput (saturation throughput of Sta-A,
the same premise keeps for future discussion) $R(T,\alpha,\mathcal{H})$
as the number of bits can be successfully transmitted per Wi-Fi slot
time. The Wi-Fi service time $D(T,\alpha,\mathcal{H})$, or the medium
access delay, is defined to be the time interval (also normalized
to system slot time) from the time instant that a packet becomes the
head of the queue and starts to contend for transmission, to the time
instant that either the packet is acknowledged for a successful transmission
or the packet is dropped. Note both $R(T,\alpha,\mathcal{H})$ and
$D(T,\alpha,\mathcal{H})$ are random variables according to above
definition. Finally, we define the average Wi-Fi throughput and average
service time $\mathcal{R}(T,\alpha,\mathcal{H})$ (in terms of bits/Wi-Fi
slot time) and $\mathcal{D}(T,\alpha,\mathcal{H})$ (in terms of Wi-Fi
time slots) respectively for a coexistence system with Wi-Fi sub-system
$\mathcal{H}(q,L,n)$ and air time sharing scheme $(T,\alpha)$ as
\begin{alignat}{1}
\mathcal{R}(T,\alpha,\mathcal{H}) & =\text{E}[R(T,\alpha,\mathcal{H})]\label{eq:Def-Throughput}\\
\mathcal{D}(T,\alpha,\mathcal{H}) & =\text{E}[D(T,\alpha,\mathcal{H})]\label{eq:Def-Service-Time}
\end{alignat}

For convenience, we sometimes omit the underlying variables $(T,\alpha,\mathcal{H})$,
and just use a simple notation as letter $R$ or $\mathcal{R}=\text{E}[R]$
for short.

\subsection{Definition of fairness}

It is a critical task to define what fairness means in this context,
since there could be many ways to describe the fairness in such a
coexistence scenario. One straightforward way is to compare the Wi-Fi
performance with and without the presence of LTE-U. More specifically,
we want to find answer to the question: Will the performance loss
(throughput degradation and service time increase) due to time sharing
be proportional to the duty cycle $\alpha$? Also, what reference
values should be used when we characterize the performance loss? The
definition below gives an intuitive way of measuring fairness.
\begin{defn}
For a given $\mathcal{H}(q,L,n)$, assume the reference Wi-Fi performance
to be $\mathcal{R}(\infty,0,\mathcal{H})$ and $\mathcal{D}(\infty,0,\mathcal{H})$.
The throughput fairness $\phi_{R}(T,\alpha,\mathcal{H})$ is the difference
between the average throughput loss ratio and the LTE-U duty cycle
$\alpha$, i.e. 
\begin{equation}
\phi_{R}(T,\alpha,\mathcal{H})=\frac{\mathcal{R}(\infty,0,\mathcal{H})-\mathcal{R}(T,\alpha,\mathcal{H})}{\mathcal{R}(\infty,0,\mathcal{H})}-\alpha\label{eq:Def-Throughput-Fairness}
\end{equation}
and service time fairness $\phi_{D}(T,\alpha,\mathcal{H})$ is the
difference of average service time increase ratio to $\frac{\alpha}{1-\alpha}$,
i.e. 
\begin{equation}
\phi_{D}(T,\alpha,\mathcal{H})=\frac{\mathcal{D}(T,\alpha,\mathcal{H})-\mathcal{D}(\infty,0,\mathcal{H})}{\mathcal{D}(\infty,0,\mathcal{H})}-\frac{\alpha}{1-\alpha}\label{eq:Def-Service-Time-Fairness}
\end{equation}
\end{defn}
Depending on $\mathcal{H}$ and $(T,\alpha)$, the fairness measures
$\phi_{R}$ and $\phi_{D}$ can be negative, positive or zero. If
both these two parameters ($\phi_{R}$ and $\phi_{D}$) are zero,
Wi-Fi performs at exact $(1-\alpha)$ ``portion'' of the non-LTE
duty cycle performance. We consider such a time sharing scheme to
be acceptable and reasonable. From this perspective, we have the following
definition. 
\begin{defn}
\label{def:Def-Fairness}A Wi-Fi LTE-U coexistence system with Wi-Fi
sub-system $\mathcal{H}$ and air time sharing scheme $(T,\alpha)$
is\textit{ fair in throughput} if $\phi_{R}\leq0$ and \textit{fair
in service time} if $\phi_{D}\leq0$. If a scheme is both fair in
throughput and service time, the scheme is \textit{fair}.
\end{defn}
\begin{rem}
Please note $q$ is a function of LTE-U to Wi-Fi INR and Wi-Fi SNR.
Formulating the collision probability $q$ is out of the scope of
this paper. It is obvious that very low INR/SNR interference causes
almost no impact to Wi-Fi system which is trivial i.e. $q\thickapprox0$
and $\phi_{R},\phi_{D}\leq0$. This paper focus on the situations
when INR and SNR are comparable, and in most subsequent discussions
we assume $q=1$, which means a Wi-Fi transmission will definitely
fail if an LTE-U transmission occurs at the same time. Additionally,
we will show in Section IV how $\phi_{D}$ and $\phi_{R}$ decay when
$q$ increases from 1 to 0 in a numerical example. Readers are reminded
that $q=1$ can happen to either strong or weak interference cases.
\end{rem}

\section{Impact of Duty Cycled LTE-U Interference}

During the LTE-U ON period, the $i$-th attempt of Wi-Fi transmission
fails with probability 
\begin{equation}
p_{i}^{'}=1-(1-p_{c})(1-q)\label{eq:LteOnpi}
\end{equation}

Whether the failure probability should be chosen as $p_{i}$ or $p_{i}^{'}$
depends on whether the LTE-U is ON or OFF, therefore the Sta-A DCF
could no longer be modeled by Markov chain. Instead, we characterize
the Sta-A throughput and service time in three steps: 
\begin{enumerate}
\item Suppose a Wi-Fi packet leaves Sta-A buffer at time $T_{0}=t_{0}$,
where $t_{0}\in\{0,1,\cdots,T-1\}$, and at time $T_{e}$, where $T_{e}\in\{t_{0},\cdots,\infty\}$,
the packet will either be sent out successfully or dropped. Conditioning
on $t_{0}$, we compute the conditional distribution $p_{T_{e}|T_{0}}(t_{e}|t_{0})$
of the finish time $T_{e}$, the conditional mean service time $\text{E}[D|t_{0}]$
and conditional mean throughput $\text{E}[R|t_{0}]$; 
\item Let $T_{e}^{'}$ be a function of $T_{e}$ that 
\[
T_{e}^{'}=T_{e}\thinspace\text{mod}\thinspace(T-1)
\]
the conditional distribution of $T_{e}^{'}$ can be derived from $p_{T_{e}|T_{0}}(t_{e}|t_{0})$
as 
\[
p_{T_{e}^{'}|T_{0}}(t_{e}^{'}|t_{0})=\sum_{T_{e}:T_{e}\thinspace\text{mod}\thinspace(T-1)=t_{e}}^{+\infty}p_{T_{e}|T_{0}}(t_{e}|t_{0})
\]
If we regard the $T$ time slots (labeled as 0,$\cdots$,$T-1$) in
a duty cycle period as $T$ states, those $T$ states form the state
space of a one dimension Markov chain, with transition probability
from state $t_{0}$ to state $t_{e}^{'}$ of 
\[
p_{T_{e}^{'}|T_{0}}(t_{e}^{'}|t_{0})
\]
because knowing the start time $t_{0}$, the distribution of $T_{e}^{'}$
does not depend on previous packet transmission start times. The distribution
$p_{T_{0}}(t_{0})$ can be computed as the stationary distribution
over these $T$ states; 
\item Lastly, the Sta-A mean throughput E$[R]$ and mean service time E$[D]$
can be derived as $\text{E}[R]=\sum_{t_{0}=0}^{T-1}p_{T_{0}}(t_{0})\text{E}[R|t_{0}]$
and $\text{E}[D]=\sum_{t_{0}=0}^{T-1}p_{T_{0}}(t_{0})\text{E}[D|t_{0}]$. 
\end{enumerate}

\subsection{The conditional probability \texorpdfstring{$p_{T_{e}|T_{0}}(t_{e}|t_{0})$}{p(te\textbar{}t0)}
under weak LTE-U interference}

When LTE-U interference is weak, Sta-A keeps transmitting in the LTE-U
ON stage whenever possible. For some $t_{0}$, consider an $m$ dimension
vector $\mathbf{w}^{(m)}=(w_{0},\cdots,w_{m-1})\in\{0,\cdots,\text{CW}_{0}-1\}\times\cdots\times\{0,\cdots,\text{CW}_{m-1}-1\}$,
where $1\leq m\leq M+1$. A vector $\mathbf{w}^{m}$ with $1\leq m<M+1$,
fully determines a back-off pattern and also uniquely determines the
transmission finish time $t_{e}$. When $m=M+1$, depending on the
last trial being successful or not, the finish time has two possibilities.
Hence the distribution $p_{T_{e}|T_{0}}(t_{e}|t_{0})$ can be derived
by first getting the conditional distribution $p_{\mathbf{W}^{(m)}|T_{0}}(\mathbf{w}^{(m)}|t_{0})$,
where $\mathbf{W}^{(m)}$ is the corresponding random variable to
$\mathbf{w}^{(m)}$.

The time of $i$-th transmission attempt $\zeta_{i}$ is also a function
of $\mathbf{w}$ that,

\begin{align}
\zeta_{i}(\mathbf{w}^{(m)})=t_{0}+\text{E}[T_{d}]\left(\sum_{k=0}^{i}w_{k}+iT_{c}\right) & \thinspace\thinspace0\leq i\leq m-1\label{eq:zetaweak}
\end{align}
compare $\zeta_{i}$ with the values of $\alpha T$, $T$, $(1+\alpha)T$,
$2T$, $\cdots$, it could be figured out if the $i$-th attempt falls
within the LTE-U ON period, record the result by a bool function 
\begin{equation}
g(\zeta_{i})=\begin{cases}
0 & \alpha T\leq\zeta_{i}\thinspace\text{mod}\thinspace T<(\zeta_{i}+T_{s})\thinspace\text{mod}\thinspace T<T\\
1 & \text{o.w.}
\end{cases}\label{eq:gi}
\end{equation}
combining \eqref{eq:LteOnpi} and \eqref{eq:gi}, we have 
\begin{align}
p(\mathbf{w}^{(m)}|t_{0}) & =\left(\prod_{i=0}^{m-2}\frac{1-(1-p_{c})\left(1-qg[\zeta_{i}(\mathbf{w}^{(m)})]\right)}{\text{CW}_{i}}\right)\nonumber \\
 & \thinspace\thinspace\thinspace\thinspace\thinspace\thinspace\thinspace\thinspace\thinspace\thinspace\thinspace\thinspace\times\left(\frac{(1-p_{c})\left(1-qg[\zeta_{m-1}(\mathbf{w}^{(m)})]\right)}{\text{CW}_{m-1}}\right)\label{eq:pWT0}
\end{align}

The end time $T_{e}$ is a function of random vector $\mathbf{W}^{(m)}$,
$T_{e}(\mathbf{W}^{(m)})$ is the sum of the total waiting time and
the time each transmission of back-off pattern $\mathbf{W}^{(m)}$.
Let $h(\mathbf{W}^{(m)})$ be a function of random variable $\mathbf{W}^{(m)}$
of the successful re-transmission probability of the $m$-th retrial,
\begin{equation}
h(\mathbf{W}^{(m)})=(1-p_{c})(1-qg(\zeta_{m-1}(\mathbf{W}^{(m)})))\label{eq:hW}
\end{equation}
According to the DCF, the mapping $f:\thinspace(\mathbf{W}^{(m)},t_{0})\rightarrow T_{e}$
is $f(\mathbf{W}^{(m)},t_{0})=$ 
\begin{equation}
\begin{cases}
t_{0}+\text{E}[T_{d}]\zeta_{m-1}+T_{s} & 1\leq m<M+1\\
t_{0}+h(\mathbf{W}^{(M)})\text{E}[T_{d}]\\
\thinspace\thinspace\thinspace\thinspace\thinspace\thinspace\times\left(\sum_{i=0}^{M}W_{i}+MT_{c}+T_{s}\right) & m=M+1\\
\thinspace\thinspace\thinspace\thinspace\thinspace\thinspace+(1-h(\mathbf{W}^{(M)}))\text{E}[T_{d}]\\
\thinspace\thinspace\thinspace\thinspace\thinspace\thinspace\times\left(\sum_{i=0}^{M}W_{i}+(M+1)T_{c}\right)
\end{cases}\label{eq:Te}
\end{equation}
Note in \eqref{eq:Te}, to make $f(\mathbf{W}^{(m)},t_{0})$ a map,
we have to deal with the map between $(\mathbf{W}^{(M+1)},t_{0})\rightarrow T_{e}$
so it has unique image, we take the last re-transmission duration
to be its expectation. The finish time $T_{e}$ has pmf 
\begin{equation}
p_{T_{e}|T_{0}}(t_{e}|t_{0})=\sum_{\mathbf{w}:f(\mathbf{w})=t_{e}}p_{\mathbf{W}|T_{0}}(\mathbf{w}|t_{0})\label{eq:pTeT0}
\end{equation}
Knowing the distribution of $T_{e}$, the conditional mean E$(D|t_{0})$
can be written as 
\begin{equation}
\text{E}[D|t_{0}]=\text{E}[T_{e}-t_{0}]\label{eq:ED_t0}
\end{equation}
For $\text{E}[R|t_{0}]$, we need to find out the probability that
a packet transmission started at $t_{0}$ being dropped, denoted as
$p_{dr}(t_{0})$, 
\begin{align}
p_{dr}(t_{0}) & =\sum_{\mathbf{w}^{(M+1)}}p_{\mathbf{W}|T_{0}}(\mathbf{w}^{(M+1)}|t_{0})[1-h(\mathbf{w}^{(M+1)})]\label{eq:pdr}
\end{align}
the conditional mean throughput can be written as

\begin{equation}
\text{E}[R|t_{0}]=L(1-p_{dr}(t_{0}))\text{E}[\frac{1}{D}|t_{0}]\label{eq:ER_t0}
\end{equation}

\subsection{The conditional probability \texorpdfstring{$p_{T_{e}|T_{0}}(t_{e}|t_{0})$}{p(te\textbar{}t0)}
under strong LTE-U interference}

Under strong interference, the Wi-Fi back-off timer will be blocked
when LTE-U is ON. As will be demonstrated later, it effectively helps
Wi-Fi to eliminate the LTE-U interference. The LTE-U interference
not only cause the collision to Wi-Fi, but also the unit decrement
time $T_{d}$ at Wi-Fi part will have a time variant pmf. When LTE-U
is ON, the Wi-Fi mean unit decrement time E$[T_{d}^{'}]$ become $\alpha T$\footnote{When $\alpha T$ is chosen at a reasonable value, for instance $T\geq100$.}
. As a result, equation \eqref{eq:zetaweak} and \eqref{eq:Te} which
both reply on E$[T_{d}]$ no longer hold true. Computation of $\zeta_{i}(\mathbf{w})$
and $T_{e}=f(\mathbf{W}^{(m)},t_{0})$ needs iterative algorithm,
which is given below, this algorithm has complexity $O(n^{2})$. Shortly
speaking, the counting process keeps checking if $T_{e}\thinspace\text{mod}\thinspace T<\alpha T$
in every iteration, if it is true then a constant $(\alpha T-S\thinspace\text{mod}\thinspace T)$
is added to the partial sum of $T_{e}$. 
\begin{align*}
 & T_{e}=t_{0}\\
 & \text{for}\thinspace i=0\thinspace:\thinspace m-1\\
 & \thinspace\thinspace\thinspace\thinspace\thinspace\thinspace\thinspace\thinspace\text{for}\thinspace j=0\thinspace:\thinspace w_{i}-1\\
 & \text{\thinspace\thinspace\thinspace\thinspace\thinspace\thinspace\thinspace\thinspace\thinspace\thinspace\thinspace\thinspace\thinspace\thinspace\thinspace\thinspace if}\thinspace(T_{e}\mod T<\alpha T)\\
 & \thinspace\thinspace\thinspace\thinspace\thinspace\thinspace\thinspace\thinspace\thinspace\thinspace\thinspace\thinspace\thinspace\thinspace\thinspace\thinspace\thinspace\thinspace\thinspace\thinspace\thinspace\thinspace\thinspace\thinspace T_{e}=T_{e}+\alpha T-T_{e}\mod T\\
 & \thinspace\thinspace\thinspace\thinspace\thinspace\thinspace\thinspace\thinspace\thinspace\thinspace\thinspace\thinspace\thinspace\thinspace\thinspace\thinspace\text{else}\\
 & \thinspace\thinspace\thinspace\thinspace\thinspace\thinspace\thinspace\thinspace\thinspace\thinspace\thinspace\thinspace\thinspace\thinspace\thinspace\thinspace\thinspace\thinspace\thinspace\thinspace\thinspace\thinspace\thinspace\thinspace T_{e}=T_{e}+\text{E}[T_{d}]\\
 & \text{\thinspace\thinspace\thinspace\thinspace\thinspace\thinspace\thinspace\thinspace\thinspace\thinspace\thinspace\thinspace\thinspace\thinspace\thinspace\thinspace}\text{end}\\
 & \thinspace\thinspace\thinspace\thinspace\thinspace\thinspace\thinspace\thinspace\text{end}\\
 & \thinspace\thinspace\thinspace\thinspace\thinspace\thinspace\thinspace\thinspace\zeta_{i}=T_{e};\\
 & \thinspace\thinspace\thinspace\thinspace\thinspace\thinspace\thinspace\thinspace\text{if}\thinspace(i\thinspace<\thinspace m-1)\\
 & \thinspace\thinspace\thinspace\thinspace\thinspace\thinspace\thinspace\thinspace\thinspace\thinspace\thinspace\thinspace\thinspace\thinspace\thinspace\thinspace T_{e}=T_{e}+T_{c}\\
 & \thinspace\thinspace\thinspace\thinspace\thinspace\thinspace\thinspace\thinspace\text{elseif}\thinspace(m\neq M+1)\\
 & \thinspace\thinspace\thinspace\thinspace\thinspace\thinspace\thinspace\thinspace\thinspace\thinspace\thinspace\thinspace\thinspace\thinspace\thinspace\thinspace T_{e}=T_{e}+T_{s};\\
 & \thinspace\thinspace\thinspace\thinspace\thinspace\thinspace\thinspace\thinspace\text{elseif}\thinspace(m=M+1)\\
 & \thinspace\thinspace\thinspace\thinspace\thinspace\thinspace\thinspace\thinspace\thinspace\thinspace\thinspace\thinspace\thinspace\thinspace\thinspace\thinspace T_{e}=T_{e}+(1-h(\mathbf{w}))T_{c}+h(\mathbf{w})T_{s}\\
 & \thinspace\thinspace\thinspace\thinspace\thinspace\thinspace\thinspace\thinspace\text{end}\\
 & \text{end}
\end{align*}

\section{Fairness Evaluation \textendash{} Monte Carlo Analysis}

It is computationally unpractical to characterize the distribution
of $p_{D}(\cdot|t_{0})$ as a function of $p(\mathbf{W}|t_{0})$ in
closed form, since the sample space of random vector $\mathbf{W}_{i}^{(M)}$
has cardinality of $\Pi_{i=0}^{M-1}\text{CW}_{i}$, which is at the
scale of $10^{11}$. However, based on the analytical discussion in
the previous section, it is easy to implement a Monte Carlo analysis
for both the weak and strong interference cases. The essential idea
of Monte Carlo analysis, also named as Monte Carlo simulation, is
to use repeated random sampling to obtain numerical results and analyze
problems that might be deterministic in principle. The procedure in
our evaluation are stepped as following, 
\begin{enumerate}
\item Let the first packet to be generated at time 0 (slot time), and at
the same time, LTE-U starts its first duty cycle; 
\item For each packet transmission/re-transmission, which may be affected
by collision (either Wi-Fi to Wi-Fi or LTE-U to Wi-Fi, or both) and
therefore multiple transmission trials may occur during one transmission/re-transmission,
we uniformly generate a sample vector $\mathbf{w}=\{w_{0},\cdots,w_{M-1}\}$
from its sample space $\prod_{i=0}^{M-1}\{0,\cdots,\mathrm{CW}_{i}-1\}$
and then a boolean vector $\boldsymbol{\epsilon}=\{\epsilon_{0},\cdots,\epsilon_{M-1}\}$
indicating success/failure for each trial whose distribution depends
on 1) The value of $\mathbf{w}$ which determines if the $i$-th Wi-Fi
transmission trial, $i\in\{0\cdots,M-1\}$, will be overlapping with
active LTE-U transmission; 2) Wi-Fi to Wi-Fi collision probability
$p_{c}$; 3) LTE-U to Wi-Fi collision probability $q$; 
\item Based on the given $\mathbf{w}$ and $\boldsymbol{\epsilon}$, we
can \textit{deterministically} compute the service time of the $k$-th
transmission and record it; 
\item Let the next transmission start immediately (because we are evaluating
saturate performance), and repeat steps 2-3 until we derive enough
numerical results for analysis; 
\item The average service time $\text{E}[D]$ as well as the throughput
$\text{E}[R]$ can be approximated using recorded service time from
each transmission. 
\end{enumerate}
We refer average throughput and average service time simply as throughput
and service time. The impact of LTE-U duty cycle is discussed in terms
of the following parameters: 
\begin{itemize}
\item duty cycle period $T$; 
\item the duty cycle $\alpha$; 
\item LTE-U to Wi-Fi collision probability $q$; 
\item Wi-Fi payload length $L$. 
\end{itemize}
On the other hand, the parameters below are fixed: 
\begin{itemize}
\item Wi-Fi system is saturated, i.e. $\lambda=0$; 
\item RTS/CTS is applied, slot time $\sigma=9$us, Wi-Fi physical layer
bit rate is 1Mb/s, for the random back-off, $M$=6, $\text{CW}_{0}=16$,
other parameters respect to IEEE 802.11n standard in the 5GHz band; 
\item The scenario contains $17$ Wi-Fi client stations, according to \cite{zhai2004performance},
we know the Wi-Fi to Wi-Fi collision probability $p_{c}=0.3739$. 
\end{itemize}

\subsection{The role of duty cycle period \texorpdfstring{$T$}{T}}

\begin{figure*}
\subfloat[throughput fairness vs $T$]{\begin{centering}
\includegraphics[scale=0.62]{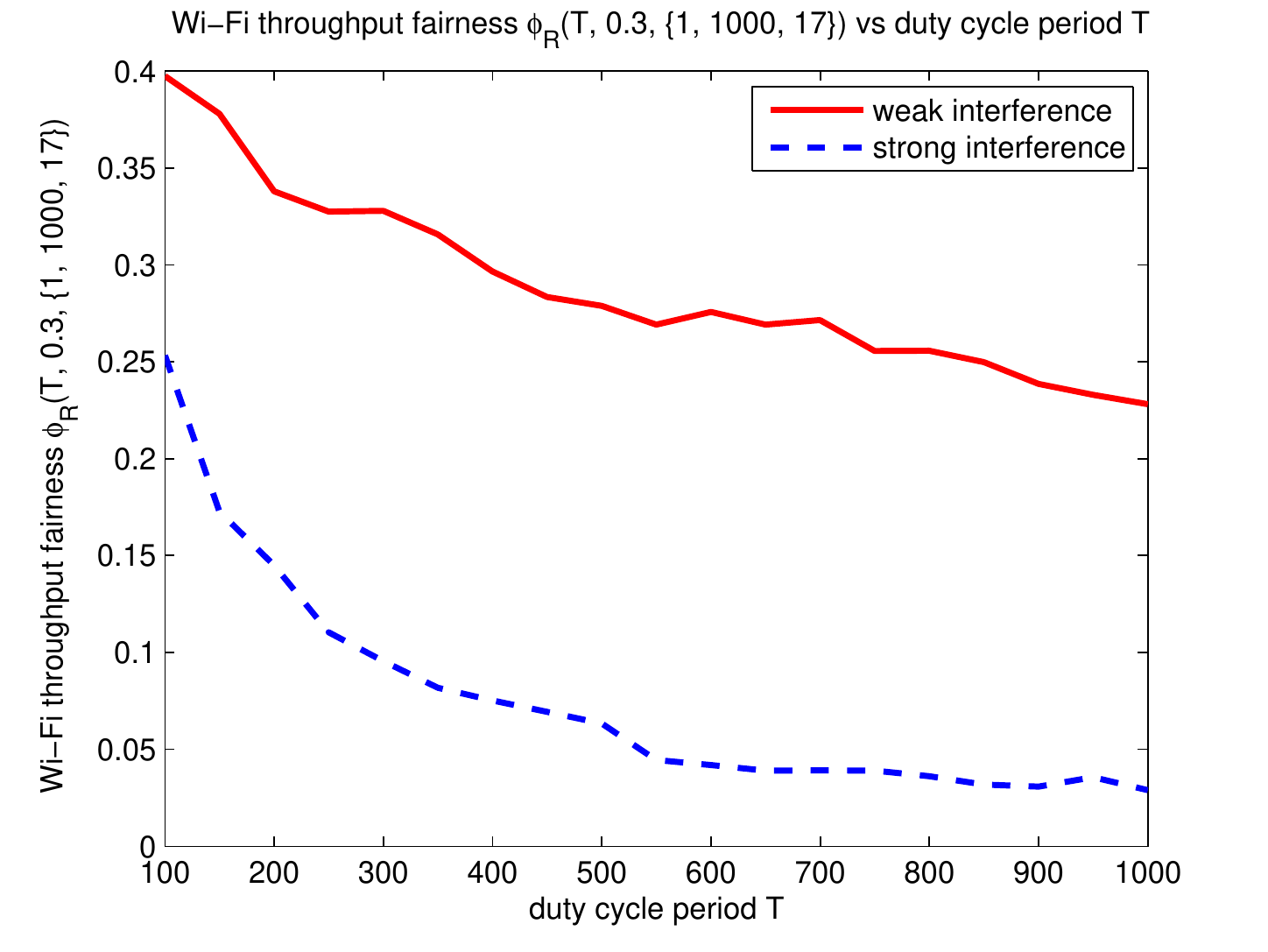} 
\par\end{centering}
}

\subfloat[service time fairness vs $T$]{\begin{centering}
\includegraphics[scale=0.62]{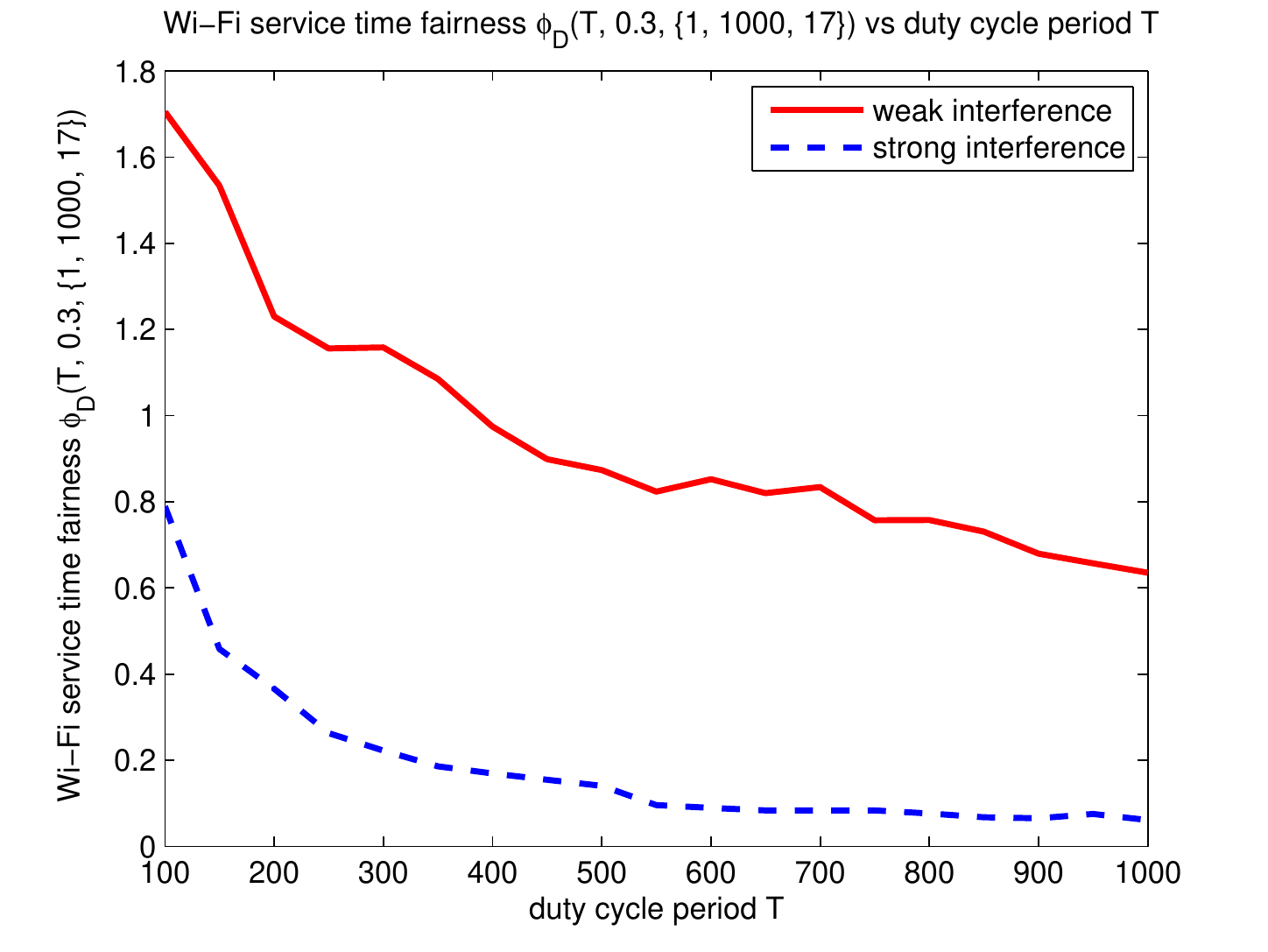} 
\par\end{centering}
}

\caption{Impact of duty cycle period $T$\label{fig:T}}
\end{figure*}

Fixing the duty cycle $\alpha=0.3$, $q=1$ and $L=1$KB, fig.\,\ref{fig:T}
shows the impact of duty cycle period $T$ on throughput and service
time, for both weak and strong interference scenarios. It can be observed
that $T\leq600$ms (note one LTE-U frame duration is 10ms) will cause
significant Wi-Fi performance degradation which is unfair to Wi-Fi.
When $T$ is large enough, the air time sharing tends to cause less
unfairness to Wi-Fi under strong interference, more numerical results
shows (omitted due the page limit) large $T$ cause less unfairness
under weak interference as well.

\subsection{The role of duty cycle \texorpdfstring{$\alpha$}{alpha}}

\begin{figure*}
\subfloat[throughput fairness vs $\alpha$]{\begin{centering}
\includegraphics[scale=0.62]{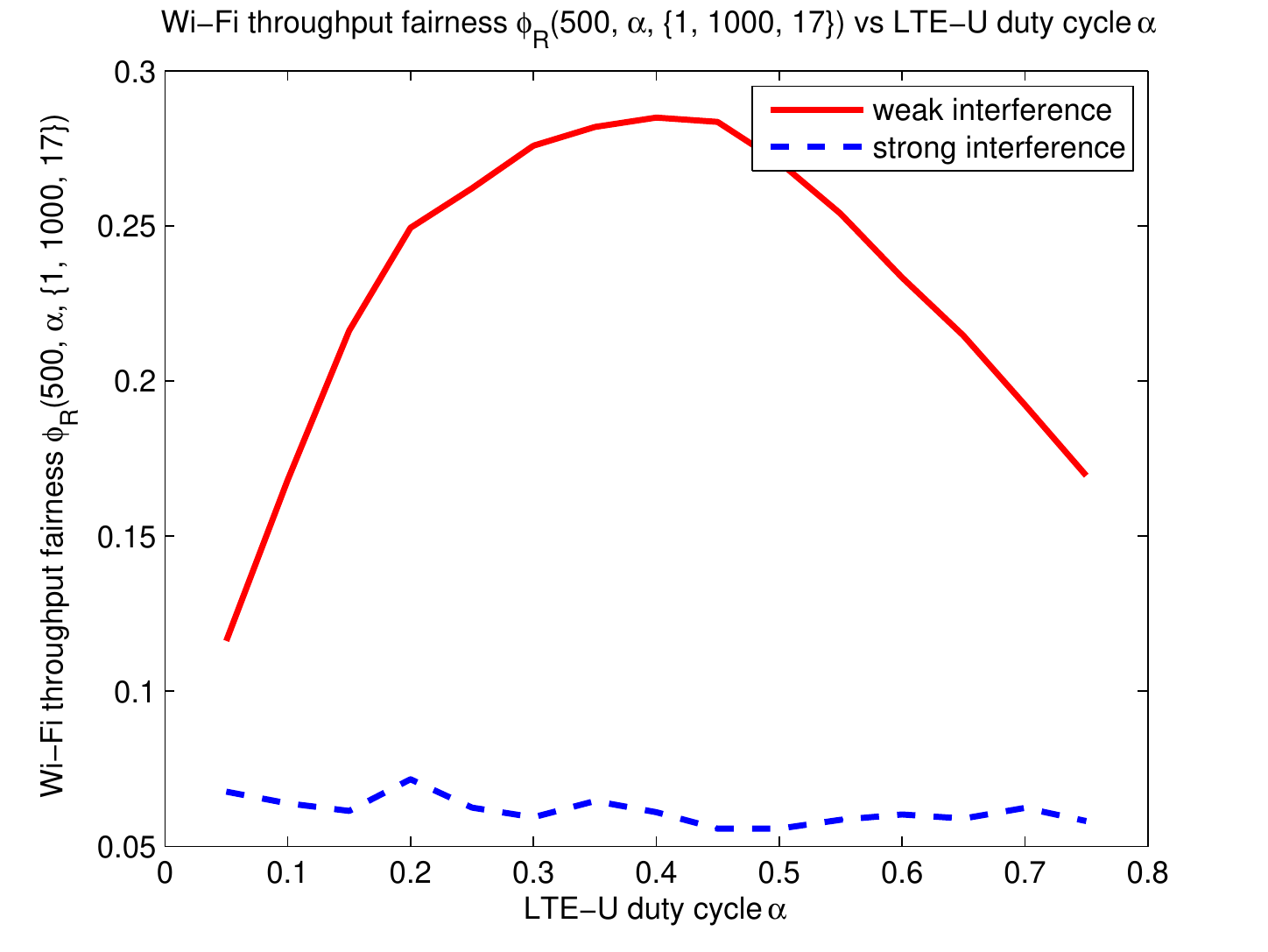} 
\par\end{centering}
}

\subfloat[service time fairness vs $\alpha$]{\begin{centering}
\includegraphics[scale=0.62]{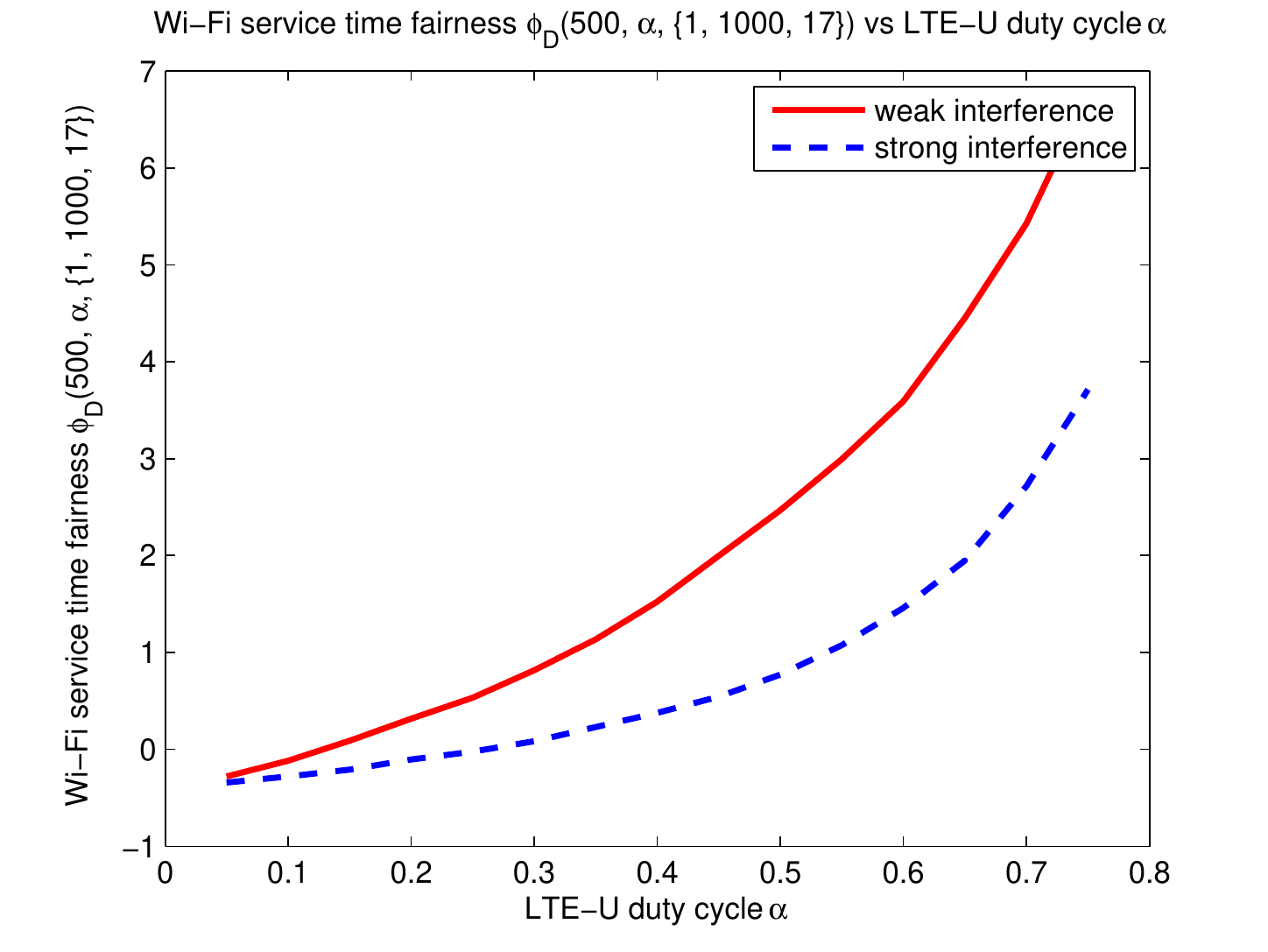} 
\par\end{centering}
}

\caption{Impact of duty cycle $\alpha$\label{fig:alpha}}
\end{figure*}

For $T=500$ms, $q=1$, and $L=1$KB, the result is demonstrated in
Fig.\,\ref{fig:alpha}. When the interference is strong, throughput
loss ratio is almost $\alpha$ and is linear. However, if interference
is weak but significant ($q$= 1), there can be additional reduction
on throughput. When $\alpha\thickapprox0.4$, the air time sharing
cause greatest throughput unfairness. When $\alpha\leq0.3,$ the delay
seems to be linear, however, when $\alpha\rightarrow1$, the service
time increase exponentially. Recall the definition of fairness and
it can be inferred for in the given network setting, a fair air time
sharing scheme should have at least $\alpha\apprle0.3$. Considering
the throughput only, strong interference approaches the throughput
fairness over almost any $\alpha\in[0,1]$.

\subsection{LTE-U to Wi-Fi collision probability \texorpdfstring{$q$}{q}}

\begin{figure*}
\subfloat[throughput fairness vs $q$]{\begin{centering}
\includegraphics[scale=0.62]{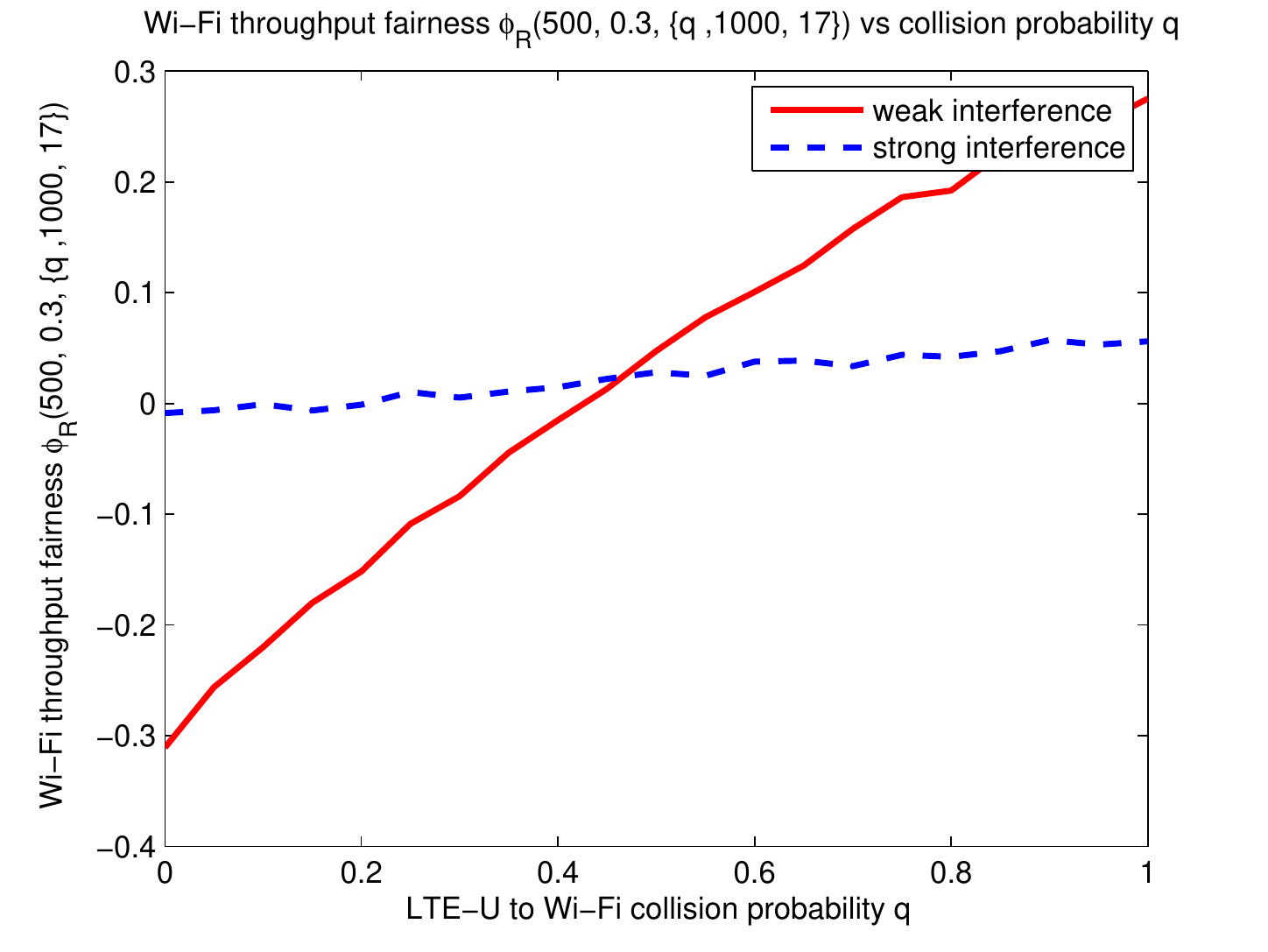} 
\par\end{centering}
}

\subfloat[service time fairness vs $q$]{\begin{centering}
\includegraphics[scale=0.62]{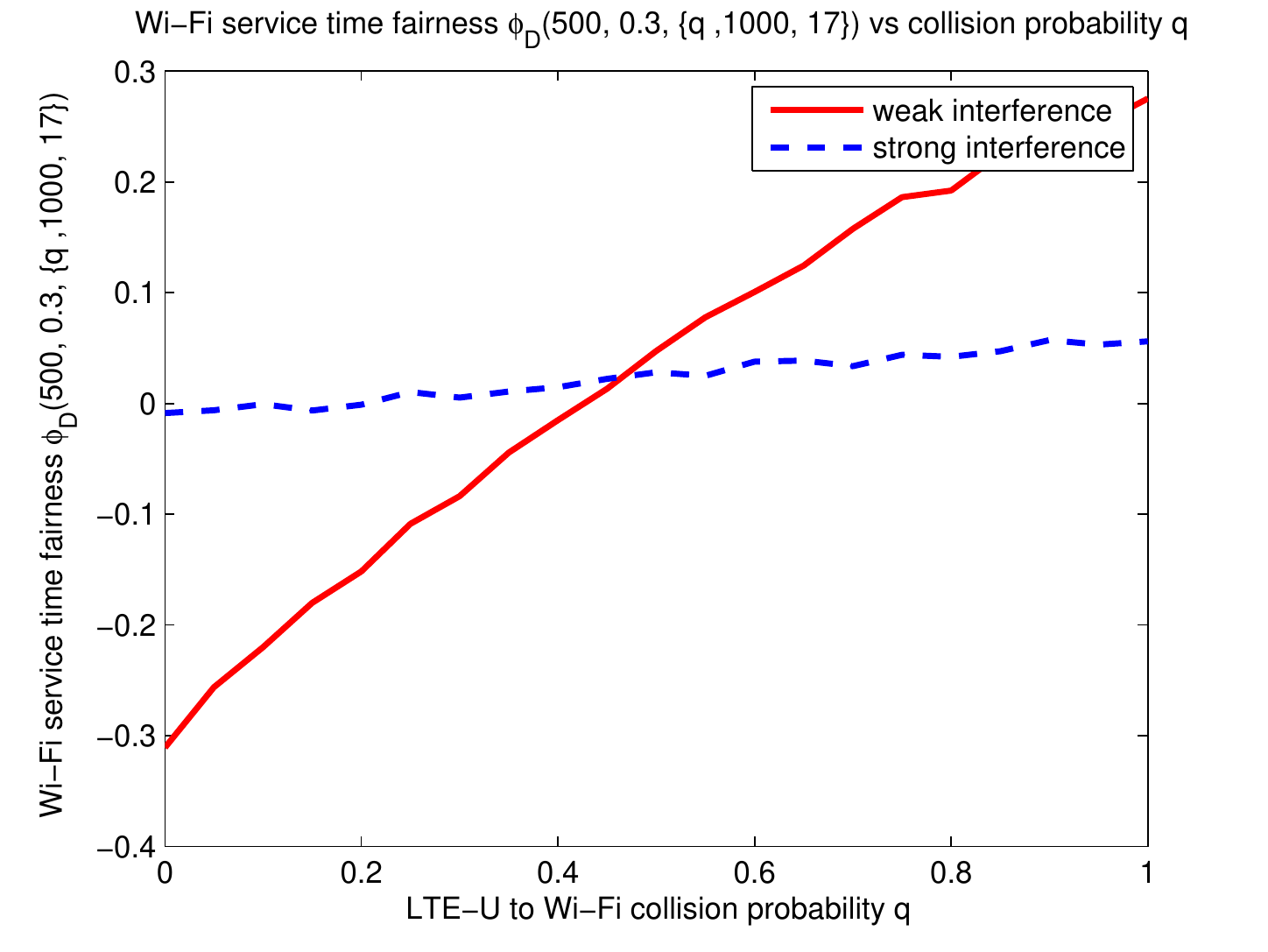} 
\par\end{centering}
}

\caption{The role of LTE-U to Wi-Fi collision probability $q$\label{fig:q}}
\end{figure*}

Fixing the duty cycle $T=500$ms and $\alpha=0.3$ and $L=1$KB, Fig.\,\ref{fig:q}
shows how the fairness varies with $q$, it degrades almost linearly
with $q$. Particularly, $q$ has less effect to fairness under strong
interference because interference over -62dbm will freeze the Wi-Fi
back-up timer and the only possible LTE-U to Wi-Fi collision occurrence
is when an LTE-U transmission starts after a Wi-Fi transmission.

\subsection{The role of Wi-Fi payload length \texorpdfstring{$L$}{L}}

\begin{figure*}
\subfloat[throughput fairness vs $L$]{\begin{centering}
\includegraphics[scale=0.62]{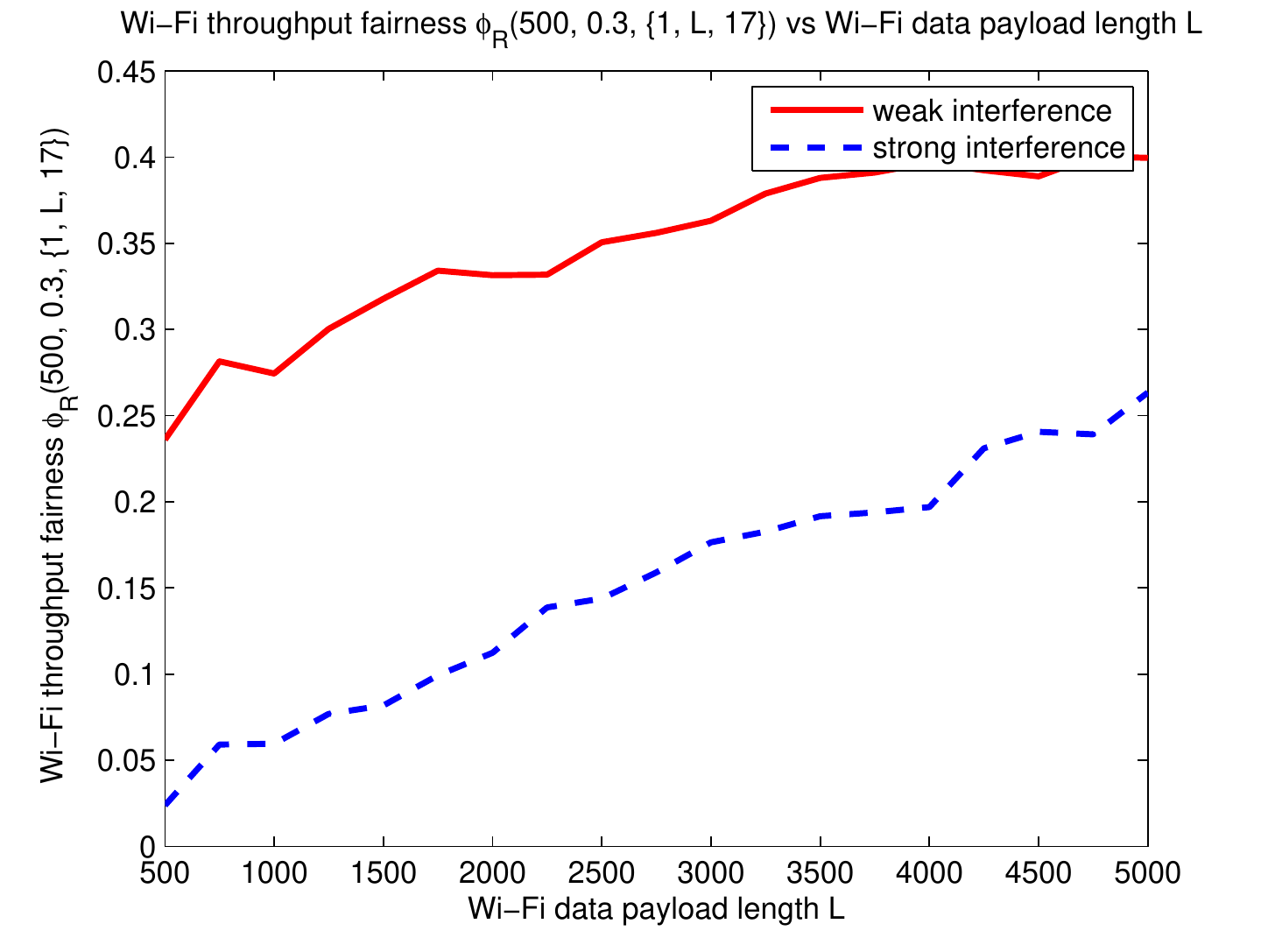} 
\par\end{centering}
}

\subfloat[service time fairness vs $L$]{\begin{centering}
\includegraphics[scale=0.62]{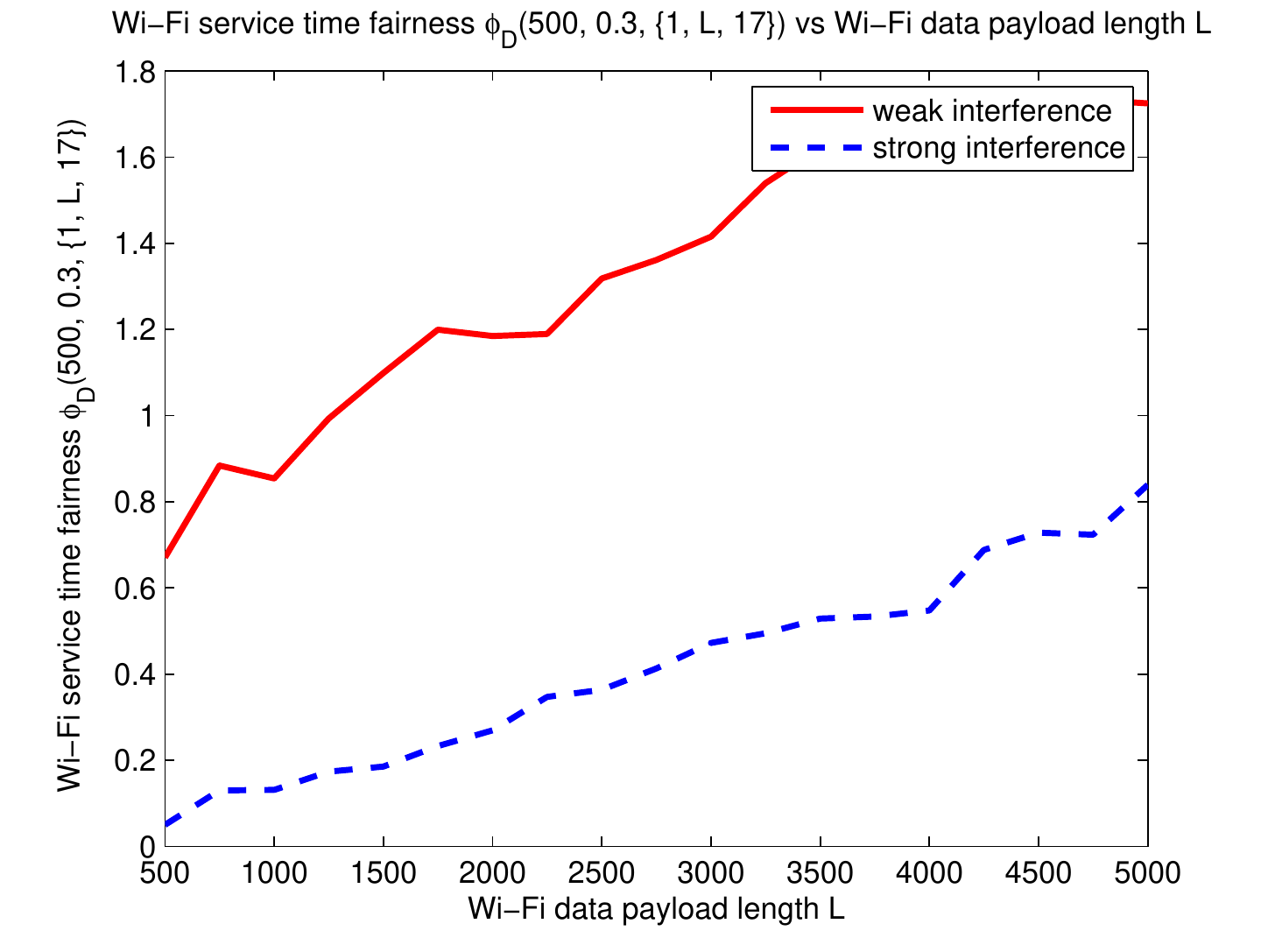} 
\par\end{centering}
}

\caption{Impact of Wi-Fi payload length $L$\label{fig:Ldat}}
\end{figure*}

Fixing $T=500$ms, $q=1$, and $\alpha=0.3$, Fig.\,\ref{fig:Ldat}
illustrates the impact of data length $L$, for both weak and strong
interference, fairness degrades almost linearly with $L$.

\subsection{Why weak interference is worse at \texorpdfstring{$q=1$}{q=1}?}

In the strong interference case, Sta-A eliminates interference by
freezing the back-off timer, after LTE-U being off, Sta-A could be
immediately released from LTE-U interference, so the total loss ratio
of the performance is very close to the the duty cycle $\alpha$,
as can be seen in Fig.\,\ref{fig:alpha}. However, when interference
is weak, Sta-A continues to contend the channel, and it eliminates
the interference by enlarging the contention window size and increasing
the number of attempts during the LTE-U-ON stage. Once LTE-U switches
OFF, Sta-A will not start transmitting until the current back-off
timer being reset, in the worst case, the recovery time could take
as long as 
\[
2^{6}\text{CW}_{0}\text{E}[T_{d}]\approx1024\mbox{E}[T_{d}]
\]
E$[T_{d}]$ in the above experiments is about 2.6ms. In case $T$
is about 100ms, Sta-A would wait for 10 duty cycle periods long before
sensing the channel again. This effect is demonstrated clearly in
Fig.\,\ref{fig:T} and \ref{fig:alpha}.

From the information theory perspective, when LTE-U interference is
strong, Wi-Fi has accurate and updated channel state information (CSI)
on whether LTE-U is on or off, hence Wi-Fi could use this CSI to skip
the interference. When LTE-U interference is weak, Wi-Fi has very
delayed CSI since the AP knows the interference only after detecting
the failure of a previous transmission, which causes significant performance
degradation.

\section{Conclusion}

In this paper, we study the performance of an infrastructure-based
Wi-Fi network when its operating channel in the unlicensed spectrum
is air time shared with an LTE-U network. We define and characterize
the Wi-Fi average performance and fairness in the presence of duty
cycled LTE-U as functions of Wi-Fi sub-system parameters and the air
time sharing scheme being used. Through Monte Carlo analysis, we numerically
demonstrate the fairness under typical coexistence settings. It can
be observed from the results that Wi-Fi and LTE-U coexistence using
simple air time sharing is generally unfair to Wi-Fi. We conclude
that some other schemes (e.g., similar to the listen-before-talk mechanism
used in 802.11 networks) need to be developed for LTE-U networks in
order to overcome the observed unfairness.

\bibliographystyle{IEEEtran}
\bibliography{Documents/CLRef}

\begin{thebibliography}{10}
\providecommand{\url}[1]{#1}
\csname url@samestyle\endcsname
\providecommand{\newblock}{\relax}
\providecommand{\bibinfo}[2]{#2}
\providecommand{\BIBentrySTDinterwordspacing}{\spaceskip=0pt\relax}
\providecommand{\BIBentryALTinterwordstretchfactor}{4}
\providecommand{\BIBentryALTinterwordspacing}{\spaceskip=\fontdimen2\font plus
\BIBentryALTinterwordstretchfactor\fontdimen3\font minus
  \fontdimen4\font\relax}
\providecommand{\BIBforeignlanguage}[2]{{%
\expandafter\ifx\csname l@#1\endcsname\relax
\typeout{** WARNING: IEEEtran.bst: No hyphenation pattern has been}%
\typeout{** loaded for the language `#1'. Using the pattern for}%
\typeout{** the default language instead.}%
\else
\language=\csname l@#1\endcsname
\fi
#2}}
\providecommand{\BIBdecl}{\relax}
\BIBdecl

\bibitem{qualcommlteuwhitepaper}
``{Extending the benefits of LTE-A to unlicensed spectrum},'' \emph{Qualcomm
  Whitepaper}, April 2014.

\bibitem{cavalcante2013performance}
A.~M. Cavalcante and et~al., ``{Performance evaluation of LTE and Wi-Fi
  coexistence in unlicensed bands},'' in \emph{Vehicular Technology Conference
  (VTC Spring), 2013 IEEE 77th}.\hskip 1em plus 0.5em minus 0.4em\relax IEEE,
  2013, pp. 1--6.

\bibitem{Babaei2014impact}
A.~Babaei, J.~Andreoli-Fang, and B.~Hamzeh, ``{On the impact of LTE-U on Wi-Fi
  performance},'' in \emph{Proceedings of IEEE PIMRC}, 2014.

\bibitem{huaweilteuwhitepaper}
``{U-LTE: Unlicensed spectrum utilization of LTE},'' \emph{Huawei Whitepaper}.

\bibitem{cano2015coexistence}
C.~Cano and D.~J. Leith, ``Coexistence of wifi and lte in unlicensed bands: A
  proportional fair allocation scheme,'' in \emph{2015 IEEE International
  Conference on Communication Workshop (ICCW)}.\hskip 1em plus 0.5em minus
  0.4em\relax IEEE, 2015, pp. 2288--2293.

\bibitem{lteuforum2015lteu}
``{LTE-U SDL Coexistence Specifications},'' \emph{LTE-U Forum}.

\bibitem{3GPPTR36889}
``{Technical Specification Group Radio Access Network; Study on
  Licensed-Assisted Access to Unlicensed Spectrum; (Release 13)},'' \emph{3GPP
  TR 36.889.0.4.0}.

\bibitem{ratasuk2014lte}
R.~Ratasuk, N.~Mangalvedhe, and A.~Ghosh, ``Lte in unlicensed spectrum using
  licensed-assisted access,'' in \emph{2014 IEEE Globecom Workshops (GC
  Wkshps)}.\hskip 1em plus 0.5em minus 0.4em\relax IEEE, 2014, pp. 746--751.

\bibitem{voicu2015coexistence}
A.~M. Voicu, L.~Simi{\'c}, and M.~Petrova, ``Coexistence of pico-and
  femto-cellular lte-unlicensed with legacy indoor wi-fi deployments,'' in
  \emph{2015 IEEE International Conference on Communication Workshop
  (ICCW)}.\hskip 1em plus 0.5em minus 0.4em\relax IEEE, 2015, pp. 2294--2300.

\bibitem{kwan2015fair}
R.~Kwan, R.~Pazhyannur, J.~Seymour, V.~Chandrasekhar, S.~Saunders, D.~Bevan,
  H.~Osman, J.~Bradford, J.~Robson, and K.~Konstantinou, ``Fair co-existence of
  licensed assisted access lte (laa-lte) and wi-fi in unlicensed spectrum,'' in
  \emph{Computer Science and Electronic Engineering Conference (CEEC), 2015
  7th}.\hskip 1em plus 0.5em minus 0.4em\relax IEEE, 2015, pp. 13--18.

\bibitem{bianchi2000performance}
G.~Bianchi, ``{Performance analysis of the IEEE 802.11 distributed coordination
  function},'' \emph{Selected Areas in Communications, IEEE Journal on},
  vol.~18, no.~3, pp. 535--547, 2000.

\bibitem{zhai2004performance}
H.~Zhai, Y.~Kwon, and Y.~Fang, ``{Performance analysis of IEEE 802.11 MAC
  protocols in wireless LANs},'' \emph{Wireless communications and mobile
  computing}, vol.~4, no.~8, pp. 917--931, 2004.

\bibitem{banchs2006end}
A.~Banchs, P.~Serrano, and A.~Azcorra, ``{End-to-end delay analysis and
  admission control in 802.11 DCF WLANs},'' \emph{Computer Communications},
  vol.~29, no.~7, pp. 842--854, 2006.

\end{thebibliography}

\end{document}